\documentclass[conference]{IEEEtran}
\IEEEoverridecommandlockouts
\usepackage{cite}
\usepackage[numbers,sort&compress]{natbib}
\usepackage{amsmath,amssymb,amsfonts}
\usepackage{algorithmic}
\usepackage{graphicx}
\usepackage{textcomp}
\usepackage{xcolor}
\usepackage{hyperref}
\usepackage{multirow}
\usepackage{booktabs}
\usepackage{ulem}
\usepackage[symbol]{footmisc}
\def\BibTeX{{\rm B\kern-.05em{\sc i\kern-.025em b}\kern-.08em
    T\kern-.1667em\lower.7ex\hbox{E}\kern-.125emX}}

\makeatletter
\newcommand{\linebreakand}{%
\end{@IEEEauthorhalign}
\hfill\mbox{}\par
\mbox{}\hfill\begin{@IEEEauthorhalign}
}
\makeatother

\IEEEpubid{\begin{minipage}{\textwidth}\ \centering
        © 2025 IEEE. Personal use of this material is permitted. Permission from IEEE must be obtained for all other uses, in any current or future media, including reprinting/republishing this material for advertising or promotional purposes, creating new collective works, for resale or redistribution to servers or lists, or reuse of any copyrighted component of this work in other works.
\end{minipage}}

\begin{document}

\title{StellarTTS: Sparse Temporal Embedding for Low-Latency and Robust Speech Synthesis\\
{\footnotesize \textsuperscript{}}
}


\author{
    \IEEEauthorblockN{Kaicheng Luo\textsuperscript{1}, Xuefei Gong\textsuperscript{1}, Yutao Sun\textsuperscript{1}, Jinling He\textsuperscript{1}, Yujie Hou\textsuperscript{1}, \\ Xiaoyang Xing\textsuperscript{1}, Huiyan Li\textsuperscript{1}, Bing Han\textsuperscript{2}, Yanmin Qian\textsuperscript{2,$\dagger$}
    }
    
    \IEEEauthorblockA{\textsuperscript{1}Honor Device Co., Ltd., China \\
    \textsuperscript{2}Auditory Cognition and Computational Acoustics Lab \\
    MoE Key Lab of Artificial Intelligence, AI Institute \\
    School of Computer Science, Shanghai Jiao Tong University, Shanghai, China\\
    \{luokaicheng1, gongxuefei\}@honor.com, \{yanminqian\}@sjtu.edu.cn
    }
}


\maketitle

\begin{abstract}
The trade-off between robustness, latency, and prosody critically challenges text-to-speech (TTS) systems. Autoregressive models, despite fidelity, are slow and error-prone; non-autoregressive (NAR) alternatives, while fast, often sacrifice prosodic naturalness via rigid alignments. This paper introduces StellarTTS, a novel mobile-optimized NAR TTS framework based on a sparse temporal embedding strategy, enabling granular control of phoneme duration, pronunciation, and prosody. Furthermore, we propose a semantic-aware codec that facilitates efficient single-stage decoding. Conditioned on the sparse temporal embedding, our 83M-parameter lightweight masked generative transformer achieves a real-time factor (RTF) of 0.08. Experiments demonstrate that StellarTTS attains lower latency and stronger robustness compared to state-of-the-art TTS systems, while maintaining competitive performance in audio quality, prosodic naturalness, and speaker similarity\footnotemark[1].
\end{abstract}

\begin{IEEEkeywords}
Zero-shot TTS, non-autoregressive models, sparse temporal embedding, semantic-aware codec, mobile deployment.
\end{IEEEkeywords}

\section{Introduction}

\footnotetext[2]{Corresponding author.}
\footnotetext[1]{
    Audio samples are available at: \url{https://stellartts.github.io/}.
}
Recent advances in zero-shot text-to-speech (TTS) systems have leveraged large-scale datasets and model architectures, spanning autoregressive (AR) \cite{valle, cosyvoice, cosyvoice2} and non-autoregressive (NAR) systems \cite{soundstorm, fastspeech,fastspeech2,e2,f5,maskgct,naturalspeech3}. While AR methods achieve high fidelity through sequential discrete token generation, error propagation in long-form synthesis remain critical challenges. CosyVoice2 \cite{cosyvoice2} makes a significant reduction in content errors by replacing VQ with FSQ and LLM initialization. However, the inherent nature of AR models introduces fundamental latency limitations for deployment in mobile devices. Additionally, AR models lack controllability during speech synthesis, making it difficult for human to refine the bad cases encountered in real applications.

NAR systems based on flow matching \cite{voicebox} or diffusion \cite{naturalspeech2,naturalspeech3} enable parallel decoding but commonly rely on explicit text-speech alignment supervision and phone-level duration prediction, creating complex pipelines that limit prosodic diversity. SoundStorm \cite{soundstorm} pioneers masked generative audio prediction, but requires semantic tokens generated by an AR model. MaskGCT \cite{maskgct} abandons text-speech alignment supervision and phoneme-level duration by using text tokens and prompt speech tokens as the prefix. Additionally, both the semantic and the acoustic model of MaskGCT are trained with the mask generative paradigm to utilize the potential of transformer. E2/F5-TTS \cite{e2,f5} also introduce a simple, fully non-autoregressive framework that offers promising naturalness and speaker similarity. Despite the impressive synthesized results of masked generative systems, we find the robustness issues of these models as reflected in a high WER, which is unacceptable in commercial application. 
\begin{figure*}[htbp]
    \centering
	\includegraphics[width=\textwidth]{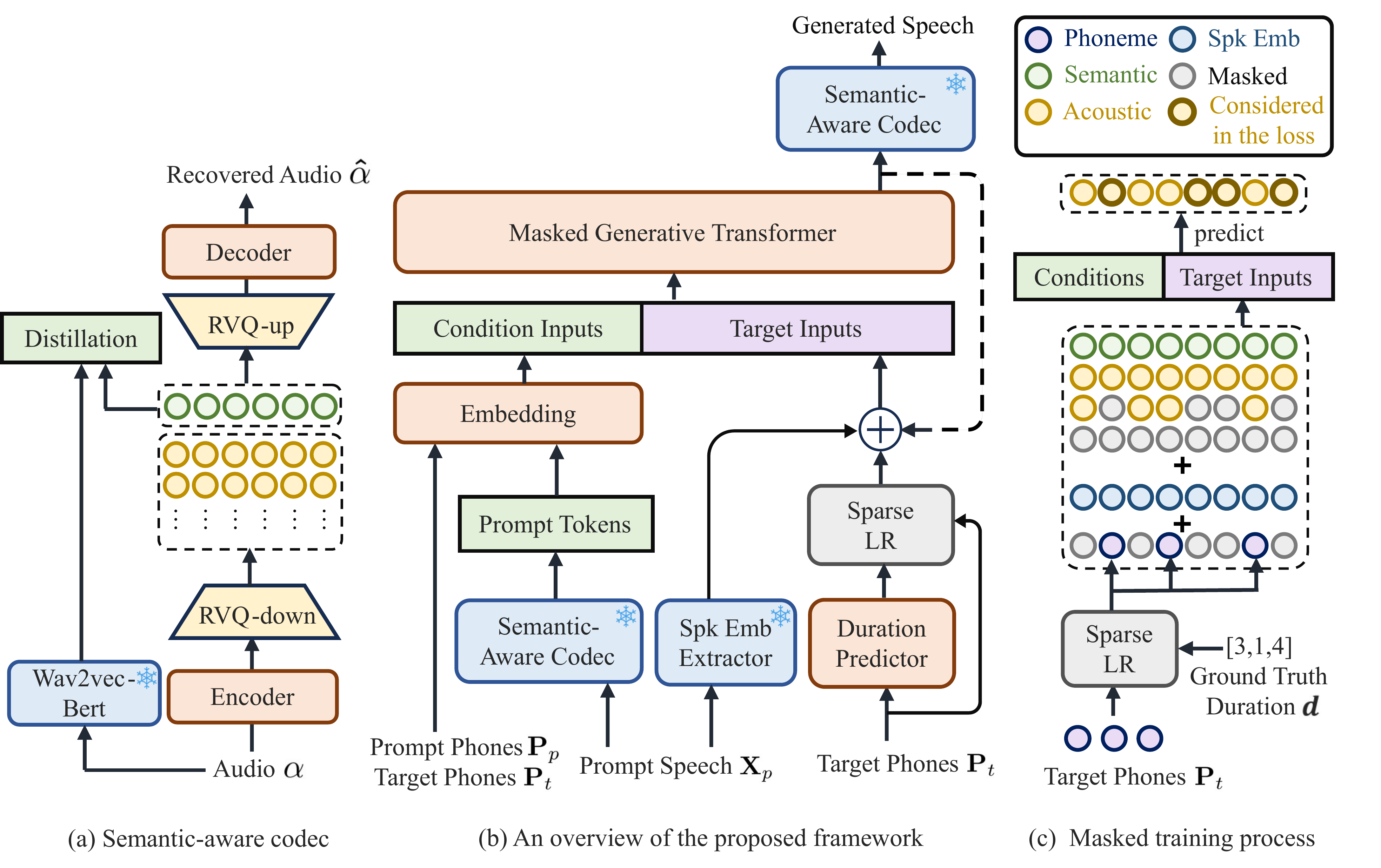}
	\caption{ An overview of StellarTTS. (a) demonstrates the semantic-aware codec, where the 0-th channel of RVQ is distillated with semantic feature extracted by Wav2vec-Bert. (b) is the overall architecture of the proposed model. The dashed lines denote that a subset of the generated tokens is selectively incorporated into the input sequence for next inference step.  (c) illustrates the training process of masked generative paradigm.}
	\label{fig1}
\end{figure*}

In this paper, we propose StellarTTS, a novel system based on \textbf{S}parse \textbf{T}emporal \textbf{E}mbedding for \textbf{L}ow-\textbf{L}atency and \textbf{R}obust \textbf{T}ext-\textbf{T}o-\textbf{S}peech synthesis. StellarTTS introduces a mobile-optimized non-autoregressive framework that significantly advances the state-of-the-art in speech synthesis by addressing critical challenges in robustness and prosodic naturalness, particularly for deployment on resource-constrained devices. Our contributions can be summarized as follows:
\begin{itemize}
\item We introduce sparse temporal embedding, a technique that significantly enhances robustness against common issues in masked generative paradigms and improves prosodic naturalness. Unlike the traditional length regulator (LR) that enforces rigid alignments, our approach allows the model to autonomously learn flexible and natural transitional patterns between phonetic units, leading to more diverse and human-like rhythms.
\item We propose a semantic-aware codec that simultaneously learns semantic and acoustic representations. This innovation simplifies the text-to-speech pipeline into a single-stage prediction process, streamlining generation and contributing to overall efficiency and model coherence. \IEEEpubidadjcol
\item We design and implement a highly efficient, mobile-first architecture centered around an 83M LLaMA-based decoder \cite{llama}.  This, combined with a lightweight backbone and single-stage decoding, enables deployment on resource-constrained platforms such as the Qualcomm SM8650. StellarTTS achieves significant improvements in inference latency (with only 16 steps) and robustness over existing systems, while maintaining competitive audio quality, prosodic richness, and speaker similarity.
\end{itemize}

\section{Related Works}

\textbf{Non-autoregressive TTS.} NAR models based on test-speech alignment \cite{fastspeech, fastspeech2} demonstrate better robustness and inference speed than AR models. NaturalSpeech3 \cite{naturalspeech3} proposes a novel framework for fine-grained speech disentanglement through a hierarchical architecture composed of diffusion-based modules. Voicebox \cite{voicebox} utilizes flow matching to perform text-guided speech infilling, allowing it to generate high-quality audio across multiple languages without the need for specialized training. SeedTTS \cite{seedtts} employs a dual-stage architecture, consisting of a phoneme-level encoder and a spectrogram decoder for efficient generation. E2/F5-TTS \cite{e2,f5} introduces a simple framework generating speech without phoneme alignment. MaskGCT \cite{maskgct} also uses prompt and text tokens as the prefix of input for mask generative transformer and achieves impressive speaker similarity.

\textbf{Discrete Speech Representation.} Discrete speech representations for zero-shot TTS have witnessed significant methodological innovations. Vector quantization approaches form the foundational paradigm, evolving from early VQ-VAE architectures \cite{discrete} to modern variants like residual-quantized SoundStream \cite{soundstream} and entropy-constrained EnCodec \cite{neuralaudio}.  Self-supervised learning techniques have expanded these representations through speech foundation models, with HuBERT-derived discrete units \cite{hubert} and WavLM-based cross-domain features \cite{wavlm} demonstrating superior speaker disentanglement. 

\section{Method}

\subsection{Semantic-Aware Codec}
Typical zero-shot TTS system separates the prediction of semantic and acoustic tokens into two stages. Early works \cite{hubert, wavlm} use k-means to quantize semantic features, which could lead to information loss. Recent papers commonly adopt VQ \cite{cosyvoice}, RVQ \cite{soundstorm, maskgct}, or FSQ \cite{cosyvoice2} to extract discrete semantic representations and demonstrate significant improvement. Inspired by \cite{soundstorm}, we propose a semantic-aware codec that employs a residual vector quantization (RVQ) framework to hierarchically quantize speech features into 1 semantic channel and 5 acoustic channels, which aims to compress the prediction of speech tokens into one stage. The training pipeline is illustrated in Fig. \ref{fig1} (a), the input audio features are sequentially processed through a multi-stage quantization pipeline, where each stage generates a residual vector for subsequent quantization.
\vspace{1em}

Specifically, the 0-th channel is optimized via knowledge distillation to preserve high-level semantic information. this is achieved by aligning its latent representation with a pre-trained semantic encoder using a KL divergence loss: 

\begin{equation}
	\begin{aligned}
	\mathcal{L}_{codec} = 
	 & \frac{1}{Td}\lambda_{rec} \cdot {\| \mathbf{\alpha - \hat{\alpha}} \|}_{2}  \\
	 &+ \frac{1}{Td}\lambda_{codebook} \cdot {\| \mathbf{sg(\varepsilon(\alpha)) - E} \|}_{2} \, \\
	 &+ \frac{1}{Td} \lambda_{commit} \cdot {\| \mathbf{sg(E) - \varepsilon(\alpha)} \|}_{2} \\
	 &+ \frac{1}{Td}\lambda_{distill} \cdot D_{KL}(f(\varepsilon(\alpha_0))\|y)
	\end{aligned}
\end{equation}

where $\alpha \in \mathbb{R}^{T\times d\times N} $ denotes an acoustic representation sequence with N channels. $\varepsilon(\alpha)$ is the output of the encoder, which is quantized to $E$ by the residual vector quantizer, and the decoder reconstructs $E$ to $\hat{\alpha}$. $ \alpha_0 $ represents 0-th channel in $\alpha$ and $y$ denotes the semantic feature extracted by Wav2vec-Bert \cite{w2v-bert}.

\subsection{Sparse Phone-level Temporal Embedding}

Classic Non-autoregressive systems use duration predictor and LR to generate temporal condition for speech prediction, which guarantees low WER while inevitably sacrificing prosody diversity. Masked generative models \cite{maskgct} without phone-level duration prediction generate more natural speech but suffer from robustness issues. We replace LR with sparse phone-level temporal embedding. During training, ground truth phone length $ t_i $ is applied to generate temporal embedding:
\begin{equation}
	\mathbf{E}_{temp} = \{\epsilon_1, \epsilon_2, ..., \epsilon_N\}
\end{equation}

\begin{equation}
	\begin{aligned}
	\epsilon_i &= \{e_1, e_2, ..., e_{t_i}\}, \, 1 <= i <= N \\
	where\ e_j &= 
		\begin{cases}
		Embed(p_i) & \text{if } j = \lfloor t_i / 2 \rfloor \\
		Embed(p_{pad}) & \text{if } j \neq \lfloor t_i / 2 \rfloor
		\end{cases}	
	\end{aligned}
\end{equation}

As shown in Fig. \ref{fig1} (c), the sparse temporal embedding strategy assigns the ground truth phone embedding $ Embed(p_i) $ exclusively to the central token position $( j = \lfloor t_i / 2 \rfloor )$ within the duration of each phone $ t_i $, while all other temporal positions are populated with a padding phone embedding $ Embed(p_{pad}) $.  The central token acts as an anchor to preserve the linguistic structure and context consistency, while the padding-dominated temporal sequence encourages the discovery of contextually adaptive prosodic variations. This balance mitigates the trade-off inherent in traditional duration-aligned methods, where explicit frame-wise alignment often results in overly deterministic prosody. 

For the absence of ground-truth phone durations during inference, we employ a duration predictor to determine both the generated audio length and the position of the central token. The duration predictor takes phone embeddings as input, optimized during training using mean squared error (MSE) loss between predicted and ground-truth phone durations. The loss function for duration prediction is formulated as:

\begin{equation}
	\mathcal{L}_{dur} = \frac{1}{N} \sum_{i=1}^{N} (\hat{d}_i - d_i)^2
\end{equation}

where $ \hat{d}_i $ denotes the predicted duration and $ d_i $ represents the ground truth duration for the $ i $-th phone. This approach ensures proper positioning of the central token at $ j = \lfloor \hat{d}_i / 2 \rfloor $ during inference while maintaining the framework's ability to generate contextually appropriate prosodic patterns.
Crucially, gradient truncation is applied to phone embeddings when serving as input to the duration predictor, preventing gradient flow from $\mathcal{L}_{dur}$ to the embedding layer. This architectural constraint ensures stable joint optimization while maintaining the linguistic integrity of the phone representations. The duration predictor receives embedding features as frozen input, but its gradient updates remain localized to the duration prediction module.

\subsection{Masked Generative Transformer}

The masked generative transformer architecture employs dual-phase attention learning with specialized embedding fusion. The input sequence $ \mathbf{X} = [\mathbf{X}_{p}; \mathbf{X}_{t}] $ undergoes differentiated masking where $ \mathbf{X}_{p} $ remains fully observable while $ X_{t} $ receives stochastic masking via Bernoulli sampling with probability $ p_{mask} $:
\begin{equation}
	where\ M_{t,c} = 
	\begin{cases}
		1, & \text{with prob } p_{mask} \\
		0, & \text{with prob } 1 - p_{mask}
	\end{cases}	
\end{equation}

where $ p_{mask} $ is dynamically modulated by the inference step index $ N_c $:
For step $ N_c = 1 $, $ p_{mask} = 1 $ (all target tokens are initially fully masked).
For step $ N_c > 1 $, $ p_{mask} $ is sampled from $ \left[ \cos\left(\frac{(N_c-1)\pi}{2N_c}\right), 1 \right] $, simulating the gradual unmasking of target tokens during the iterative inference steps. This design mimics the iterative refinement process, where higher steps $ N_c $ allow increased exposure of target tokens through the sinusoidal bound probability $ p_{mask} $.

The model input comprises two components: a condition prefix and target embeddings. The condition prefix integrates contextual cues by concatenating three discrete token sequences: prompt phones $ \mathbf{P}_p $, target phones $ \mathbf{P}_t $, and prompt acoustic tokens $ \mathbf{X}_{p} $, separated by delimiter tokens $ s_1,s_2,s_3 $. Each sequence is individually embedded through trainable lookup tables:
\begin{equation}
	\mathbf{E}_{prefix} = \left[ 
     Emb(\mathbf{P}_{p};s_1); 
	 Emb(\mathbf{P}_{t};s_2); 
	 Emb(\mathbf{X}_{p};s_3) 
	 \right]
\end{equation}

where $ Emb(\cdot) $ denotes the embedding layer and $ [\cdot; \cdot] $ represents the concatenation along the temporal dimension.

The target embeddings align temporally with the output audio tokens and combines three elements:

The target audio tokens $ \mathbf{X}_t $ are masked and embedded as: $ \mathbf{E}_{token}=Embbeding(\mathbf{X}_{t} \odot M) $. The speaker embedding extracted by the pre-trained voice-print model \cite{3d-speaker} is projected through a linear layer and expanded to align with the token embedding in the time dimension: $ \mathbf{E}_{spk}' = Expand(Linear(\mathbf{E}_{spk}))$.
These components and the sparse temporal embedding $\mathbf{E}_{temp}$ mentioned in the last subsection are summed to form the target embedding:
\begin{equation}
	\mathbf{E}_{target} = \mathbf{E}_{token}
	+ \mathbf{E}_{spk}'
	+ \mathbf{E}_{temp}
\end{equation}

As shown in Fig. \ref{fig1} (b), the final decoder input is constructed by concatenating the condition prefix and target embeddings:
\begin{equation}
\mathbf{E}_{\text{input}} = \left[ \mathbf{E}_{prefix}; \mathbf{E}_{target} \right]
\end{equation}

enabling joint conditioning on linguistic, acoustic, and speaker contexts while preserving temporal guidance for parallel token prediction.

The decoder processes the combined input representation to produce classification logits for parallel token prediction. The cross-entropy loss is computed over the masked positions as:
\begin{equation}
\mathcal{L}_{\text{CE}} = - \sum_{i=1}^N m_{i,t} \cdot \mathbf{y}_i \log (P_{\theta}(x_i|\mathbf{X}_{t} \odot M, \mathbf{C}))
\end{equation}

where $m_{i,t}$ denotes the masking indicator, $\mathbf{y}_i$ denotes the target token distribution, and $P_{\theta}$ refers to the conditional probability distribution parameterized by $\theta$. The total training objective combines both alignment and prediction losses through weighted summation:
\begin{equation}
	\mathcal{L}_{total} = \mathcal{L}_{CE} + \lambda \mathcal{L}_{dur}
\end{equation}

\section{Experiments and Results}

\textbf{Datasets.} We utilize Emilia \cite{emilia}, an in-the-wild multilingual speech dataset, to train our model. For the test set, we perform evaluation on the Seed-TTS test-zh and test-hard \cite{seedtts} set to compare our model with baselines, among which the test-hard is a very challenging set that includes sentences with especially challenging patterns such as word repetitions, tongue twisters, etc.

\textbf{Training.} Our models are trained on 8 NVIDIA A100 40GB GPUs. We use the AdamW optimizer with a peak learning rate of 5e-4, linearly warmed up for 10,000 steps, and decays during the rest of training. The weight decay is set to 0.01. During the training process, we drop the target prompt following the cosine scheduler $ p_{mask} \in  \left[ \cos\left(\frac{(N_c-1)\pi}{2N_c}\right), 1 \right]$.

\textbf{Inference.} We use inference steps [8,4,1,1,1,1] for RVQ layers. Greedy sampling is employed for the channels where the inference step is 1. For channel 0 and channel 1, we use the top 10\% logits with a temperature annealing from 0.1 to 0. Gumbel noise is adopted for token confidences that determine the tokens to be remasked. 

\textbf{Evaluation Metrics.} Following previous works \cite{f5,maskgct}, we utilize the word error rate (WER), speaker similarity between generated and original target speeches (SIM-o), and real-time factor (RTF) for objective evaluation. In detail, the WavLM-TDCNN speaker embedding model \cite{wavlm} is deployed to estimate SIM-o and a HuBERT-based ASR model \cite{hubert} is applied to calculate WER. RTF is computed on single NVIDIA A100 GPU with 10-second speech samples. Comparative mean opinion scores (CMOS) and similarity mean opinion scores (SMOS) of our model and baselines are measured for subjective evaluation.

\subsection{Objective Results}
SIM-o, WER and RTF are employed for objective assessment. We compare our model with the AR and NAR baselines.

\begin{table}[h]
	\centering
	\caption{Objective evaluation results on Seed-TTS test-zh and test-hard sets. The boldface indicates the best result.}
	\label{tab:objective}
	\begin{tabular}{lccccc}
		\toprule
		\multirow{2}{*}{\textbf{Model}} & \multicolumn{2}{c}{\textbf{test-zh }} & \multicolumn{2}{c}{\textbf{test-hard }} &  \\
		\cmidrule(lr){2-3} \cmidrule(lr){4-5}
		 & \textbf{WER}↓ & \textbf{SIM-o}↑   & \textbf{WER}↓ & \textbf{SIM-o}↑ & \textbf{RTF}↓ \\
		\midrule
		Ground Truth &  1.26 & 0.755 & - & - & - \\
		\hline
		FireRedTTS\cite{firered} &  1.51 & 0.668  & 17.45 & 0.621 & - \\
		CosyVoice\cite{cosyvoice} & 3.63 & 0.723  & 11.75 & 0.709 & 0.67 \\
		MaskGCT\cite{maskgct} & 2.27 & \textbf{0.774}  & 10.27 & \textbf{0.748} & 0.71 \\
		F5-TTS\cite{f5} & 1.56 & 0.741  & 8.67 & 0.713 & 0.31 \\
		\hline
		StellarTTS & \textbf{1.44} & 0.712  & \textbf{7.47} & 0.697 & \textbf{0.08} \\
		\bottomrule
	\end{tabular}
\end{table}

As shown in Table \ref{tab:objective}, our model demonstrates a superior performance to the state-of-the-art models in terms of WER, especially on hard cases, highlighting its enhanced robustness. Furthermore, the proposed architecture yields a $4\times\sim9\times$ acceleration in inference speed compared to baselines, facilitating efficient edge deployment with constrained resources. 

While our model exhibits lower performance on the SIM-o metric compared to state-of-the-art models, this discrepancy stems from the design of speech representation. Specifically, we conducted comparative analyzes between the vocoder reconstruction of real speech, revealing that our codec underperforms others in the SIM-o metric (0.668). The performance gap could be attributed to semantic distillation during vocoder training, which is intended to compress the inference pipeline into a single stage. This trade-off aligns with our objective of prioritizing real-time edge deployment without compromising overall robustness.

\begin{table}[h]
	\centering
	\caption{Subjective evaluation results on Seed-TTS test-zh and test-hard sets. (mean scores)}
	\label{tab:subjective}
	\begin{tabular}{lccccc}
		\toprule
		\multirow{2}{*}{\textbf{Model}} & \multicolumn{2}{c}{\textbf{test-zh }} & \multicolumn{2}{c}{\textbf{test-hard }} \\
		\cmidrule(lr){2-3} \cmidrule(lr){4-5}
		& \textbf{CMOS}↑ & \textbf{SMOS}↑ & \textbf{CMOS}↑ & \textbf{SMOS}↑ \\
		\midrule
		Ground Truth & 0.00 & 3.86 & 0.00 & 3.85 \\
		\hline
		FireRedTTS\cite{firered} & -0.49 & 3.28 & -0.76 & 3.24 \\
		CosyVoice\cite{cosyvoice} & -0.14 & 3.54 & -0.22 & 3.48 \\
		MaskGCT\cite{maskgct} & -0.08 &\textbf{4.09} & -0.12 & \textbf{3.86} \\
		F5-TTS\cite{f5} &  0.02 & 3.83 & -0.01 & 3.72 \\
		\hline
		StellarTTS & \textbf{0.06} & 3.96 & \textbf{0.05} & 3.78 \\
		\bottomrule
	\end{tabular}
\end{table}

\subsection{Subjective Results}
We conduct human assessments with 20 native speakers for CMOS and SMOS, presenting paired comparisons in random order. The result is shown in Table \ref{tab:subjective}. The result demonstrates that StellarTTS outperforms other baselines in speech naturalness and robustness while maintaining competitive speaker similarity. Although MaskGCT attains marginally higher SMOS, its end-to-end synthesis architecture exhibits fundamental limitations in handling linguistically complex scenarios. Specifically, the character omission errors of MaskGCT under challenging conditions degrade its perceptual naturalness, as reflected in lower CMOS. 

\begin{table}[h]
	\centering
	\caption{Ablation studies of sparse strategies and speaker embedding.}
	\label{tab:ablation}
	\begin{tabular}{lcccc}
		\toprule
		\multirow{2}{*}{\textbf{Model}} & \multicolumn{2}{c}{\textbf{test-zh}} & \multicolumn{2}{c}{\textbf{test-hard}} \\
		\cmidrule(lr){2-3} \cmidrule(lr){4-5}
		 & \textbf{WER}↓ & \textbf{SIM-o}↑ & \textbf{WER}↓ & \textbf{SIM-o}↑ \\
		\midrule 
		StellarTTS  & 1.44 & 0.712 & 7.47 & 0.697 \\
		w/o central strategy  & 2.34 & 0.701  & 9.15 & 0.688 \\
		w/o temporal emb. & 4.10  & 0.693 & 18.12 & 0.670 \\
		w/o speaker emb. & 1.58 & 0.654 & 7.31 & 0.626 \\
		\bottomrule
	\end{tabular}
\end{table}

\subsection{Ablation Study}
We explore the impact of the sparse strategy for temporal embedding and speaker embedding on WER and SIM-o. Specifically, central strategy is the proposed method where center tokens in phones are unmasked when generating sparse temporal embeddings and random strategy refers to where a token within the phone duration is randomly selected as the anchor for sparse temporal embedding. The results, summarized in Table \ref{tab:ablation}, reveal the relative contributions of sparse temporal embedding and speaker embedding to model performance. Notably, the central strategy consistently outperforms the random strategy in terms of WER, suggesting that anchoring at the temporal center provides more stable and informative cues for the model. Furthermore, the absence of sparse temporal embedding or speaker embedding leads to a critical degradation in WER and SIM-o respectively, highlighting the necessity of both components for context consistency and speaker similarity.

\subsection{Phoneme Duration Control}

\begin{figure}[htbp]
	\includegraphics[width=0.48\textwidth]{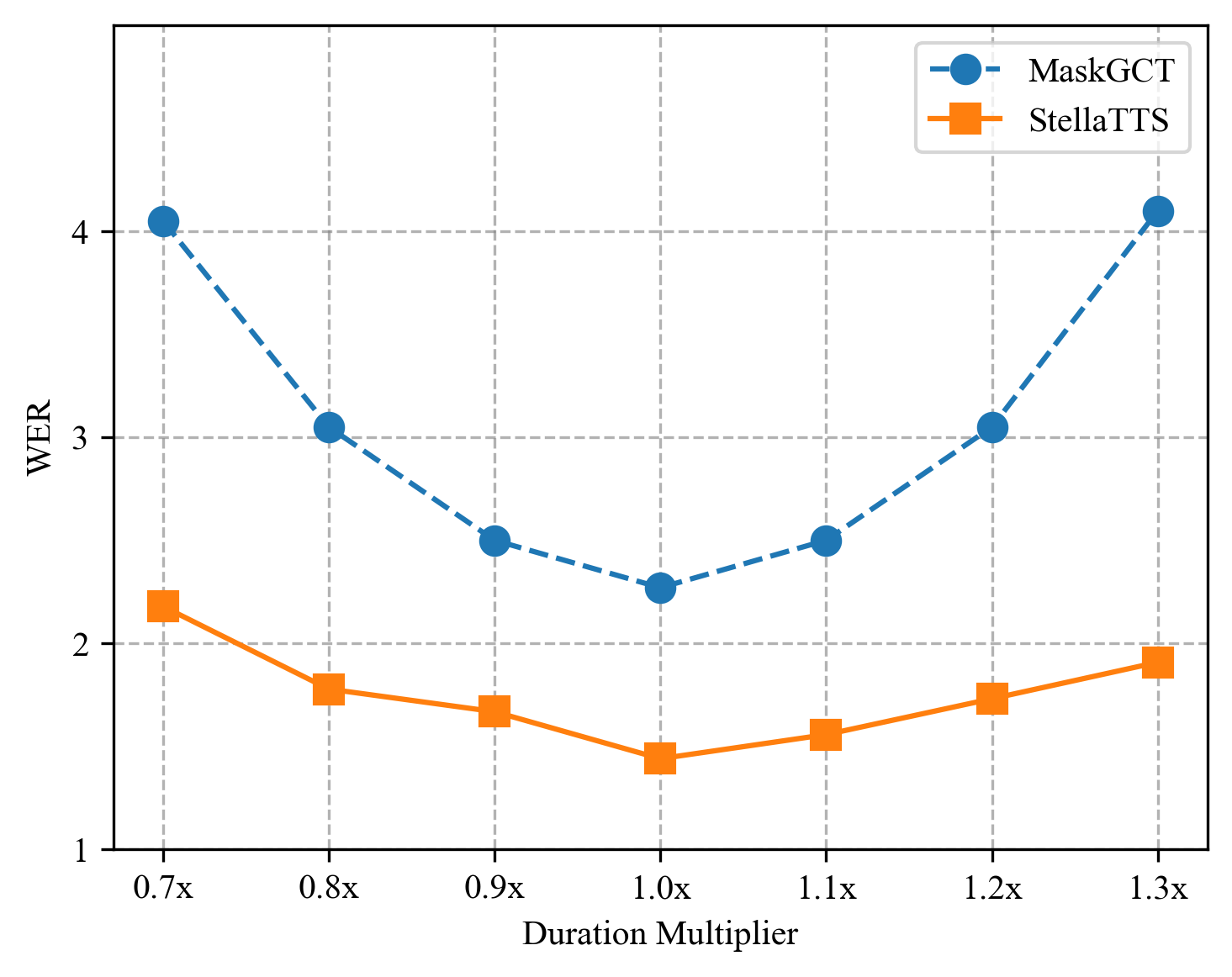}
	\caption{Impact of Duration Control Strategies on WER: StellarTTS and MaskGCT Evaluated on Seed-TTS test-zh.}
	\label{fig2}
\end{figure}

We analyze the robustness of StellarTTS in generating speech with different speeds by modifying the duration of each phoneme. Additionally, we compare the phoneme duration control strategy in our model with the total duration control strategy in MaskGCT. We evaluated speech by multiplying the optimal predicted phoneme duration by 0.7 to 1.3, which is shown in Figure \ref{fig2}. The results of MaskGCT are obtained from the paper, generated by multiplying the total duration of the ground truth by 0.7 to 1.3. The results demonstrate that the phoneme duration control in our model achieves stronger robustness among different duration lengths. When the duration multiplier is 0.7, the WER of our model is the highest (2.18), which is still lower than the best WER of MaskGCT (2.27). This showcases that our model achieves precise phoneme duration control in synthesized speech without sacrificing semantic fidelity, which facilitates human refinement on generated results.

\section{Conclusion}
This work introduces StellarTTS, a mobile-optimized non-autoregressive text-to-speech system utilizing semantic-aware codec, sparse temporal embedding, and a lightweight single-stage decoding architecture. By integrating semantic and acoustic representations within a unified codec framework, the model eliminates multi-stage prediction dependencies in existing systems. The proposed sparse temporal embedding mechanism overcomes the flat prosody produced by classic NAR methods and robustness issue in pipelines without phoneme alignment, allowing precise control of phoneme duration, pronunciation and prosody. Coupled with single-stage inference and the lightweight decoder design, the architecture achieves significant improvements in efficiency and deployment flexibility. Experiments demonstrate that StellarTTS outperforms the state-of-the-art systems in robustness and efficiency, with notable reductions in linguistic errors and accelerated inference speeds. Optimized for mobile deployment, the model balances computational cost with high-fidelity speech synthesis, showcasing practical viability for industrial applications that require real-time responsiveness and resource efficiency.

\end{document}